\documentclass[aps,pra,reprint,10pt,superscriptaddress,longbibliography,floatfix]{revtex4-2}

\usepackage{amsmath}
\usepackage{amssymb}
\usepackage{bm}
\usepackage{graphicx}
\usepackage{mathtools}
\usepackage{xcolor}
\usepackage{microtype}
\usepackage[colorlinks=true,linkcolor=blue,citecolor=blue,urlcolor=blue]{hyperref}
\hypersetup{hidelinks}
\graphicspath{{figures/}}

\newcommand{\E}{\mathbb{E}}
\providecommand{\tr}{\operatorname{tr}}
\providecommand{\ket}[1]{\lvert #1\rangle}

\newcommand{\SigSAT}{$\Sigma$-SAT}

\begin{document}

\title{Cross-chip transfer in quantum photonic neural networks}

\author{Iqra Bashir}
\affiliation{College of Physics and Optoelectronic Engineering, Shenzhen University, Shenzhen 518060, China}
\author{Bikash K. Behera}
\affiliation{Bikash's Quantum (OPC) Pvt. Ltd., Mohanpur, WB 741246, India}
\affiliation{Universit\`a degli Studi di Cagliari, Via Is Mirrions, Cagliari 09123, Italy}
\author{Qi Qin}
\email[Contact author: ]{qiqin32@cityu.edu.hk}
\affiliation{Department of System Engineering, City University of Hong Kong, 83 Tat Chee Avenue, Kowloon, Hong Kong SAR, China}
\affiliation{Quantum Science Center of Guangdong-Hong Kong-Macao Greater Bay Area, Shenzhen 518045, China}
\author{Keren Li}
\email[Contact author: ]{likr@szu.edu.cn}
\affiliation{College of Physics and Optoelectronic Engineering, Shenzhen University, Shenzhen 518060, China}
\affiliation{Quantum Science Center of Guangdong-Hong Kong-Macao Greater Bay Area, Shenzhen 518045, China}

\begin{abstract}
Fabrication variability makes it difficult to reuse one set of trained
parameters across a batch of quantum photonic chips without additional
chip-specific optimization. We examine how device-specific compensation and
task tolerance shape cross-chip performance.
For squared operator loss, local analysis predicts that reducing own-chip error
through compensation incurs an increase in expected loss on independent
chips. We use Monte Carlo ensemble training (MC-ENS), informed by an available
statistical model of fabrication-induced parameter errors. It optimizes one
shared parameter set for each task or target at a given
fabrication scale by averaging losses over simulated chip
realizations resampled from this model. Numerical simulations of a quantum photonic neural network
compare MC-ENS with single-chip training, sharpness-aware training and a
covariance-informed variant. For CNOT synthesis and Ising dynamics,
MC-ENS reduces mean conditional process infidelity and increases mean
postselection probability on unseen chips. In classification, high accuracy
can persist despite substantial changes in the optical transformation.
The decision-margin analysis shows when fabrication-induced score shifts
are likely to change predicted labels.
The trained shared parameters can be reused without
device-specific retraining on unseen chips drawn from the same fabrication
distribution.
\end{abstract}
\maketitle

\section{Introduction}
\label{sec:intro}
Programmable quantum photonic chips use trained optical controls to implement
computational tasks~\citep{steinbrecher2019quantum}. A practical deployment goal
is to obtain one parameter set for a given task or target and reuse it across
chips of the same design without device-specific retraining. Fabrication
variability complicates this transfer: chips with the same layout can differ in
coupler splitting ratios and interferometric phases, and
therefore implement different transformations
\citep{ewaniuk2023imperfect,xing2023spatial}. Such differences can degrade task
performance when the same controls are reused.

Device-specific training and calibration can compensate for fabrication
errors in individual photonic circuits
\citep{ewaniuk2023imperfect,bandyopadhyay2021hardware}. Training informed by
hardware statistics uses weight perturbations in electronic
neural-network hardware~\citep{joshi2020accurate} and signal perturbations in
silicon-photonic architectures~\citep{mourgias2022noise}. Xu \textit{et al.}
linked loss-landscape flatness to physical robustness and demonstrated
cross-device parameter transfer using sharpness-aware training (SAT)
\citep{xu2026physical}. SAT seeks low loss within local neighborhoods of the
trainable parameters. Cohen \textit{et al.} used a related
noise-aware approach to component design, optimizing fixed waveguide
geometries for photonic quantum gates using fidelity averaged over
fabrication errors~\citep{cohen2025robust}. At the system level, the challenge
is to preserve task performance when a common set of programmable controls
is reused across chips.

Nominal accuracy and robustness to systematic errors are distinct
properties of a control implementation~\citep{brown2004arbitrarily}.
For cross-chip deployment, this distinction calls for understanding how
corrections fitted to one fabrication realization affect performance on another.
The consequences also depend on whether a task requires an accurate operator,
a target quantum state or a correct class label
\citep{gilchrist2005distance,weber2021optimal}. We therefore ask how
chip-specific compensation contributes to transfer error and when
fabrication-induced output changes impair task performance.

To investigate these questions, we model static fabrication errors using
a parameter distribution that can incorporate available process statistics,
prior device characterization or engineering estimates. Monte Carlo ensemble
training (MC-ENS) resamples virtual chips from this distribution at each
optimization step in a quantum photonic
neural network (QPNN) simulator. It optimizes shared trainable controls by
minimizing the mean task loss across the sampled chips. This objective follows
the principles of domain randomization~\citep{tobin2017domain} and stochastic
programming~\citep{shapiro2009lectures}. Using per-device (PD) training and
SAT as reference methods,
we evaluate cross-chip transfer in nonunitary operator learning,
Bell-state preparation, CNOT synthesis
and Ising dynamics. A local analysis examines how chip-specific compensation
affects the expected loss on independent chips. With trained controls held fixed,
we assess how sensitive the implemented transformations are to small
fabrication errors. For operator learning, we also examine how the
chip-averaged transformation deviates from the target at larger variability.
Finally, we use sample-level decision margins in Banknote classification to
explain when predicted classes remain unchanged despite fabrication-induced
variations in optical outputs.

\section{Model and Training Objectives}
\label{sec:model}

\begin{figure}[tb]
\centering
\includegraphics[width=\columnwidth]{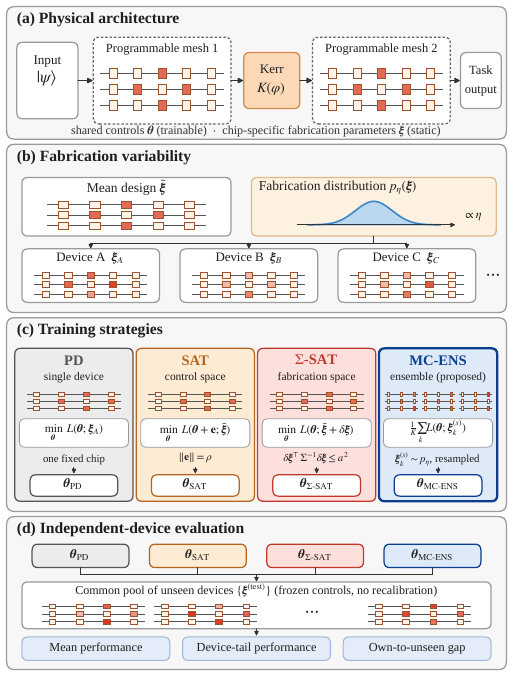}
\caption{Cross-chip transfer framework. (a) Programmable QPNN with
shared controls $\bm\theta$ and chip-specific fabrication parameters
$\bm\xi$. (b) Fabrication distribution $p_\eta$ around the mean design
$\bar{\bm\xi}$, from which individual devices are drawn.
(c) MC-ENS training and the comparison methods PD, SAT and $\Sigma$-SAT.
(d) Frozen parameters evaluated on a common pool of independently
sampled, unseen devices.}
\label{fig:framework}
\end{figure}

Architectures combining programmable linear optics with optical
nonlinearities have been applied to quantum gates, state compression and
quantum simulation~\citep{steinbrecher2019quantum,nielsen2025programmable}.
We consider a six-mode, two-photon Fock-state quantum photonic neural
network (QPNN) for tractable ensemble studies of cross-chip transfer in
quantum and classification tasks.
It comprises two programmable linear-optical meshes separated by a
Kerr nonlinear operation $K(\varphi)$
[Fig.~\ref{fig:framework}(a)].
The Kerr operation applies a phase $e^{i\varphi}$ to states with
both photons in the same mode and leaves all other two-photon states unchanged.
The meshes comprise phase shifters,
directional couplers and lossy waveguides, with overall transformation
\begin{equation}
S(\bm\theta;\bm\xi)
=
S_2(\bm\theta_2;\bm\xi_2)
K(\varphi)
S_1(\bm\theta_1;\bm\xi_1),
\label{eq:network}
\end{equation}
where $\bm\theta$ collects the programmable optical controls,
including $\varphi$, and $\bm\xi$ represents the static fabrication
parameters of a chip.
With loss included, $S$ gives the unnormalized two-photon output amplitudes.

Chip-to-chip variability is represented by a distribution
$p_\eta(\bm\xi)$ over fabrication parameters, including coupler
transmittances and propagation losses, with $\eta$ controlling the
magnitude of variability [Fig.~\ref{fig:framework}(b)].
This distribution can be informed by prior device characterization,
process statistics or engineering estimates~\citep{ewaniuk2023imperfect,xing2023spatial}.
For each chip, fabrication parameters are treated as fixed during operation.
The numerical architecture and the explicit definition of $p_\eta$,
including Gaussian sampling parameters and physical clipping, are given
in Supplemental~Material, Sec.~S1~\cite{supplemental}.

For a given task or target and fabrication scale $\eta$,
one parameter set $\bm\vartheta$,
comprising the optical controls and any trainable encoding or readout
parameters, is shared across devices drawn from $p_\eta$.
The corresponding deployment
objective is the expected task loss $
R_\eta(\bm\vartheta)
=
\mathbb E_{\bm\xi\sim p_\eta}
\left[L(\bm\vartheta;\bm\xi)\right].
$
MC-ENS$_K$ optimizes this objective in a differentiable simulator by
sampling $K$ independent device realizations at each optimization
step $s$ and averaging their separately evaluated losses:
\begin{equation}
\widehat R^{(s)}_{\eta,K}(\bm\vartheta)
=
\frac{1}{K}\sum_{k=1}^{K}
L(\bm\vartheta;\bm\xi_k^{(s)}),
\qquad
\bm\xi_k^{(s)}\overset{\mathrm{iid}}{\sim}p_\eta.
\label{eq:ensemble_loss}
\end{equation}
The device realizations are resampled at every step, including for
MC-ENS$_1$. Increasing $K$ changes the sampling variance and
evaluations per update, but not the expected objective.

We use per-device (PD) training and SAT, together with its
covariance-informed variant $\Sigma$-SAT, as reference methods
[Fig.~\ref{fig:framework}(c)]. PD training minimizes
$L(\bm\vartheta;\bm\xi_A)$ on one fixed fabricated chip, unlike
MC-ENS$_1$, which resamples at every step.
SAT~\citep{xu2026physical} is implemented on the design circuit using
first-order perturbations of the trainable parameters to seek
low-loss neighborhoods. $\Sigma$-SAT instead applies first-order
adversarial perturbations to fabrication parameters within a
covariance-shaped neighborhood. Both use gradients evaluated at
perturbed configurations rather than ensemble averaging.
Their fixed-radius candidates are reused across
$\eta$, with radii selected on a separate validation-device pool at each scale.
All methods use the same task loss, optimizer
and initialization protocol within each benchmark; detailed updates
and training settings are given in Supplemental~Material, Sec.~S1.

After training, all parameters are frozen and evaluated without
recalibration on a common pool of 1000 unseen devices at each
scale from $\eta=0.05$ to 4: $\eta\in\{0.05,0.10,\ldots,0.50\}$
and $\{1.0,1.5,\ldots,4.0\}$ for operator learning, and
$\eta\in\{0.5,1.0,\ldots,4.0\}$ for the other tasks, sampled from $p_\eta$ independently of training
[Fig.~\ref{fig:framework}(d)].
PD is also evaluated on its own device using the same inputs.
We report the device-pool mean and a device-tail statistic
(95th-percentile error or 5th-percentile accuracy).
Error bars and quoted intervals denote 95\% bootstrap confidence
intervals. See Supplemental~Material, Table~S1 and Sec.~S2, for
the numbers of targets and training runs and the evaluation metrics,
respectively.

\section{Device-specific adaptation and cross-device robustness}
\label{sec:results}

\subsection{Operator learning}
We first use nonunitary operator learning to study
cross-chip parameter transfer. The objective is to
approximate a target operator $M\in\mathbb{C}^{4\times4}$ using the
post-selected computational-subspace transformation
$T(\bm\theta;\bm\xi)$ implemented by the QPNN. To distinguish
fabrication-induced errors from limitations in target realizability,
we construct ten target operators from randomly configured mean-design
circuits. For each device, the training loss is
\[
L(\bm\theta;\bm\xi)
=
\frac{1}{16}
\left\lVert T(\bm\theta;\bm\xi)-M\right\rVert_F^2.
\]
MC-ENS$_K$ averages this loss over $K$ sampled devices at each
update [Eq.~\eqref{eq:ensemble_loss}].
We report the relative operator error
$\varepsilon_{\rm rel}=\lVert T-M\rVert_F/\lVert M\rVert_F$ and use
its square, $\ell=\varepsilon_{\rm rel}^2$, for compensation and
fabrication-response analyses.
A separate attainability check on the mean-design circuit recovers
all targets with $\varepsilon_{\rm rel}<10^{-5}$.

\begin{figure}[!htbp]
\centering
\includegraphics[width=\columnwidth]{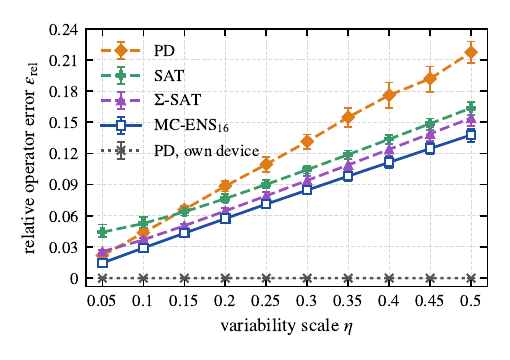}
\caption{Mean relative operator error on unseen devices versus
fabrication variability $\eta$ for PD, SAT, $\Sigma$-SAT and
MC-ENS$_{16}$. Error bars denote 95\% bootstrap confidence intervals
over targets.}
\label{fig:operator}
\end{figure}

As shown in Fig.~\ref{fig:operator}, MC-ENS$_{16}$ yields the lowest mean unseen-device operator error at every tested variability, $0.05\le\eta\le0.5$, and across all ten targets. At $\eta=0.3$, the mean relative errors are 0.131, 0.104, 0.094 and 0.085 for PD, SAT, $\Sigma$-SAT and MC-ENS$_{16}$, respectively. The reduction relative to PD remains 33--37\% throughout the tested range, while the reduction relative to SAT decreases from 67\% at $\eta=0.05$ to 16\% at $\eta=0.5$, and that relative to $\Sigma$-SAT decreases from 43\% at $\eta=0.05$ to 10--11\% for $\eta\ge0.25$. The larger reductions at small $\eta$ depend on the fixed SAT and $\Sigma$-SAT radii; with smaller radii, MC-ENS$_{16}$ remains lower by 4--12\% (Supplemental~Material, Sec.~S8). In contrast, PD achieves $\varepsilon_{\mathrm{rel}}\approx10^{-6}$ on its own training device. This disparity shows that accurate device-specific fitting
does not ensure reliable parameter transfer.

\subsection{Local compensation and transfer cost}
\label{sec:transfer-mechanisms}

To explain the PD transfer gap, we linearize the operator
residual about a solution that is exact on the mean-design circuit.
For the squared loss $\ell=\varepsilon_{\rm rel}^2$, let
$\bar\ell_0\propto\eta^2$ be the expected loss of these fixed controls
on fabricated chips and $f\in[0,1]$ the fraction of this leading-order
loss that local control adjustments can compensate.

\begin{figure}[!htbp]
\centering
\includegraphics[width=\columnwidth]{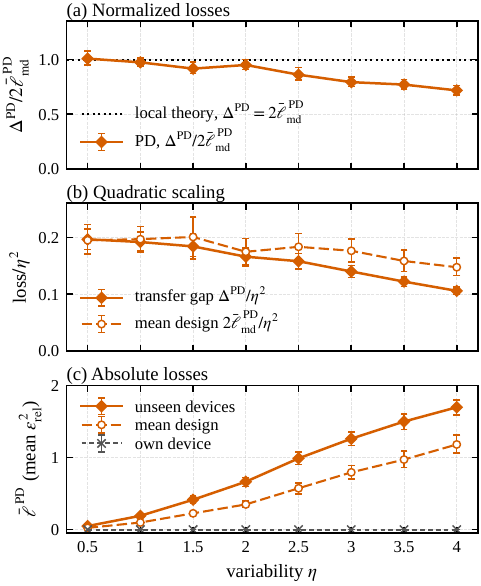}
\caption{Local compensation mechanism in operator learning (PD controls only).
(a) Transfer gap $\Delta^{\rm PD}$ normalized by twice the loss of the same
controls on the mean-design circuit; the dotted line is the local-theory
prediction, Eq.~\eqref{eq:gap}. (b) Transfer gap
$\Delta^{\rm PD}$ and twice the mean-design loss
$2\bar\ell_{\rm md}^{\rm PD}$, both normalized by $\eta^2$.
(c) Corresponding squared losses
$\varepsilon_{\rm rel}^2$ before normalization, on unseen devices, on the
mean-design circuit and on the own training device. Losses are averaged over
training runs within each target; error bars are paired 95\% bootstrap
intervals over the ten targets.}
\label{fig:mechanism}
\end{figure}

For small, centered fabrication perturbations, PD controls that minimize
the linearized squared loss give the following mean losses on the training
chip (own), independent chips from the same distribution (unseen) and
the mean-design circuit (md):

\begin{equation}
\bar\ell^{\rm PD}_{\rm own}\simeq(1-f)\bar\ell_0,
\bar\ell^{\rm PD}_{\rm unseen}\simeq(1+f)\bar\ell_0,
\bar\ell^{\rm PD}_{\rm md}\simeq f\bar\ell_0.
\label{eq:local}
\end{equation}
Relative to $\bar\ell_0$, compensation subtracts $f\bar\ell_0$
from the mean own-chip loss and adds it to the mean unseen-chip loss, giving
\begin{equation}
\Delta^{\rm PD}
\equiv
\bar\ell^{\rm PD}_{\rm unseen}
-
\bar\ell^{\rm PD}_{\rm own}
\simeq
2\bar\ell^{\rm PD}_{\rm md}
\propto\eta^2.
\label{eq:gap}
\end{equation}
The mean-design loss isolates the effect of the retained
correction (Supplemental~Material, Sec.~S3).

Figure~\ref{fig:mechanism} tests Eq.~\eqref{eq:gap} for PD.
The own-chip loss is below $2\times10^{-7}$ for every PD model, so
the gap is dominated by the unseen-chip loss. At small variability,
the gap and the mean-design loss scale approximately as $\eta^2$.
At $\eta=1$, the ratio $\Delta^{\rm PD}/(2\bar\ell^{\rm PD}_{\rm md})$
is 0.97 [0.94, 1.01], consistent with the predicted value of 1.
The first tested scale whose interval excludes 1 is $\eta=1.5$
(0.92 [0.87, 0.98]), and the ratio falls to 0.72 [0.68, 0.76]
at $\eta=4$.

In the same fixed-response local model, the population optimum of
MC-ENS and the unchanged mean-design controls both yield an expected
unseen-chip squared loss of $\bar\ell_0$, which is $f\bar\ell_0$ below
that of locally optimal PD (Supplemental~Material, Sec.~S3).

\subsection{Transfer across quantum objectives}

\begin{figure}[!htbp]
\centering
\includegraphics[width=\columnwidth]{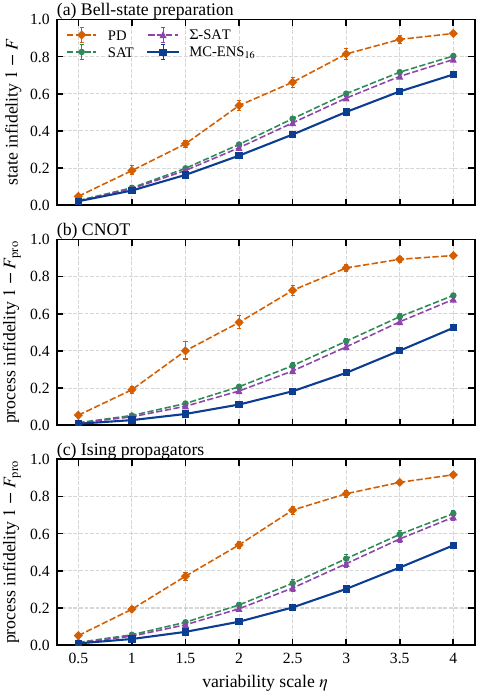}
\caption{Mean conditional infidelity on unseen chips versus $\eta$
for PD, SAT, $\Sigma$-SAT, and MC-ENS$_{16}$. (a) State infidelity for
Bell-state preparation conditioned on two-photon survival.
(b) and (c) Conditional process infidelity for CNOT and Ising dynamics,
respectively. Error bars denote 95\% bootstrap
confidence intervals over training runs (a,b) or targets (c).}
\label{fig:quantum}
\end{figure}

To test cross-chip transfer beyond operator learning, we
simulate three quantum tasks (Fig.~\ref{fig:quantum}). Bell pairs supply
entanglement for quantum teleportation~\citep{bennett1993teleporting}, and
CNOT is a building block of universal quantum circuits~\citep{barenco1995elementary}.
Two-qubit transverse-field Ising dynamics provide a benchmark for photonic
simulation of interacting spins~\citep{steinbrecher2019quantum}. Separate
optimizations on the mean-design circuit reach infidelities below $10^{-4}$
for all targets.

\emph{State preparation.}
The Bell task maps $\ket{101000}$ with vacuum ancillas to the
dual-rail state $\ket{\Phi^+}=(\ket{101000}+\ket{010100})/\sqrt{2}$.
For the unnormalized two-photon output $\ket{\psi}$, the conditional
state fidelity is
$F=|\langle\Phi^+|\psi\rangle|^2/\langle\psi|\psi\rangle$,
where the denominator is the two-photon survival probability.

\emph{CNOT implementation.}
For post-selected CNOT on two dual-rail qubits, we compare the
computational-subspace transformation $T$ with the ideal gate $G$ using
the conditional process fidelity
$F_{\rm pro}=|\operatorname{tr}(G^\dagger T)|^2/
[4\operatorname{tr}(T^\dagger T)]$.
This metric is invariant to an overall nonzero complex scale.

\emph{Ising dynamics.}
We train separate controls for ten propagators $U(t)=e^{-iHt}$
of the two-qubit transverse-field Ising Hamiltonian
$H=Z_AZ_B+h(X_A+X_B)$ and use the same conditional process fidelity.
For evolution from $\ket{00}$, we also evaluate the conditional
state fidelity and the error in $\langle Z_A\rangle$.

MC-ENS$_{16}$ yields the lowest mean conditional infidelity in all
three tasks at every tested $\eta$. Relative to $\Sigma$-SAT, it reduces
mean infidelity by 9--14\% for Bell-state preparation, 23--41\% for CNOT,
and 22--36\% for Ising dynamics. At $\eta=1$, for example, the mean
CNOT process infidelity is 0.191 for PD and 0.027 for MC-ENS$_{16}$
[Fig.~\ref{fig:quantum}(b)]. The state and magnetization diagnostics
for Ising evolution show corresponding improvements. The relative gains
for CNOT and Ising narrow at larger variability, and all four methods
have mean process infidelities above 0.5 at $\eta=4$.
See Supplemental~Material, Sec.~S4, for task definitions, auxiliary
observables, and the separately assessed survival and postselection
probabilities.

\section{Fabrication response of shared controls}
\label{sec:sensitivity}
Controls with similar mean-design accuracy can respond
differently to fabrication. For operator learning, let $r$ be the
real-vectorized residual $(T-M)/\lVert M\rVert_F$. Near the
mean-design circuit, $r\simeq r_0+J\delta\bm\theta+B\delta\bm\xi$,
where $r_0$ is the nominal residual and $J$ and $B$ are the control
and fabrication Jacobians at the chosen solution. For fixed controls
and centered fabrication perturbations with covariance $\eta^2\Sigma_0$,
the linearized model gives

\begin{equation}
\E_{\bm\xi}[\ell]\simeq\lVert r_0\rVert^2
+\eta^2\,\tr\!\left[B\Sigma_0B^{\mathsf T}\right].
\label{eq:sensitivity-risk}
\end{equation}

For $r_0\neq0$, residual curvature can also contribute at order
$\eta^2$ (Supplemental~Material, Sec.~S6).
SAT probes the response through $J$. This can help transfer when
control and fabrication perturbations produce comparable changes in
the residual. \SigSAT\ probes fabrication directly using a
covariance-informed first-order ascent direction. MC-ENS averages
the loss over fabrication realizations. At comparable nominal errors,
the average in Eq.~\eqref{eq:sensitivity-risk} favors smaller
covariance-weighted fabrication responses.

\begin{figure}[tb]
\centering
\includegraphics[width=\columnwidth]{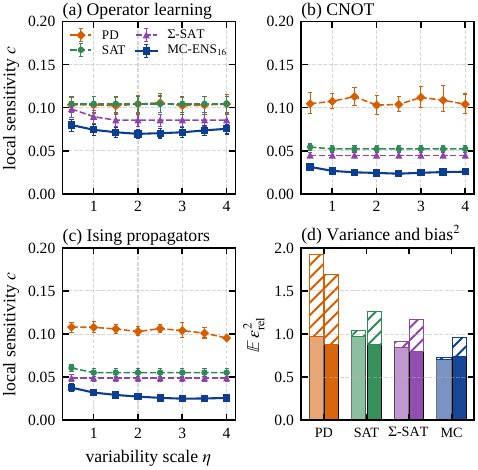}
\caption{Fabrication response of frozen controls.
(a)--(c) Local sensitivity $c$ measured at probe scale $\eta_p=0.1$
for controls trained or selected at each $\eta$ in operator learning,
CNOT, and Ising dynamics. Comparison methods are shown dashed and
MC-ENS$_{16}$ solid, and error bars denote 95\% bootstrap confidence
intervals over targets or training runs.
(d) Exact decomposition of the mean squared relative operator error on unseen chips into the variance across chips (filled) and the squared bias of the chip-averaged
transformation (hatched) at $\eta=1$ (left bars, multiplied by 10) and
$\eta=4$ (right bars).}
\label{fig:sensitivity}
\end{figure}

The local fabrication sensitivity $c$, defined in
Supplemental~Material, Sec.~S6, measures departures from each frozen
model's mean-design transformation. We estimate it on the same 500
probe chips at $\eta_p=0.1$ [Fig.~\ref{fig:sensitivity}(a)--(c)].
For operator learning, $c\simeq\tr(B\Sigma_0B^{\mathsf T})$.
CNOT and Ising use a normalized process distance.
At $\eta=1$, PD and SAT have nearly
equal sensitivities in operator learning ($c=0.103$ and 0.104),
consistent with the retained correction in Sec.~\ref{sec:transfer-mechanisms}.
\SigSAT\ and MC-ENS reduce $c$ by 14\% and 29\% relative to SAT.
For CNOT and Ising, SAT reduces $c$ by 52\% and 49\% relative to PD,
respectively, and MC-ENS reduces it by 49\% and 42\% relative to SAT.
At $\eta=1$, the estimate $E_{\rm md}+\eta^2c$, with $E_{\rm md}$
the mean-design error, tracks the variation in unseen-chip errors in
each task $R^2\ge0.97$ for the unadjusted predictions,
Supplemental~Material, Sec.~S6 and Fig.~S2).

Figure~\ref{fig:sensitivity}(d) decomposes the mean squared relative
operator error on unseen chips into the variance across chips and the
squared bias of the chip-averaged transformation at $\eta=1$ and 4
(Supplemental~Material, Sec.~S6). PD and SAT have similar variance, and
PD's excess error is mainly due to the bias term. For MC-ENS relative
to \SigSAT, lower variance accounts for 76\% of the improvement at
$\eta=1$. At $\eta=4$, the chip-averaged transformation for MC-ENS
is closer to the target, and lower squared bias accounts for 73\%
of the improvement.

In operator learning, MC-ENS$_1$ reduced transfer error at the
same budget of simulated loss-and-gradient evaluations as PD.
At comparable budgets, taking more updates with one fabrication draw
per update ($K=1$) gave lower error than using $K=16$. In the tested
correlated-error setting, misspecified fabrication correlations reduced
the transfer gains of MC-ENS and \SigSAT\ (Supplemental~Material,
Sec.~S7).

\section{Task-Dependent Robustness and Classification}
\label{sec:classification}

Predicted classes can remain unchanged as optical
outputs vary across chips. We study this effect on the Banknote
authentication dataset~\citep{lohweg2012banknote} (1372 samples,
four features, two classes). Data re-uploading encodes the features in
optical phases~\citep{perezsalinas2020data}. A trainable affine readout
maps four dual-rail detection probabilities to two class scores.
We train with cross-entropy loss.

\begin{figure}[!htbp]
\centering
\includegraphics[width=\columnwidth]{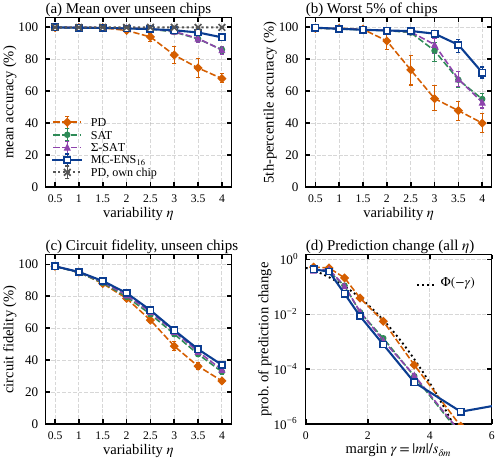}
\caption{Task-dependent robustness in Banknote classification.
(a) Mean and (b) 5th-percentile accuracy on unseen chips; the dotted
gray curve in (a) shows PD on its training chip. (c) Mean circuit
fidelity relative to the designed circuit. Error bars in (a)--(c) are 95\% bootstrap confidence intervals over training runs. (d) Probability of a prediction change relative to the mean-design circuit, binned by normalized margin $\gamma$ and pooled
over all eight values of $\eta$. The dotted curve is the zero-mean
Gaussian reference $\Phi(-\gamma)$.}
\label{fig:classification}
\end{figure}

Test accuracy is the fraction of held-out samples correctly classified on a given chip. Figures~\ref{fig:classification}(a) and (b) show the mean and 5th-percentile accuracy across unseen chips, respectively, for $\eta=0.5,1.0,\ldots,4.0$. All four methods exceed 99.5\% mean accuracy for $\eta\le1.5$. Despite achieving at least 99.9\% accuracy on its training chip at every $\eta$, PD exhibits the earliest degradation on unseen chips. The three shared-control methods maintain mean accuracies within 0.2 percentage points of one another up to $\eta=2.5$. At $\eta=3$, MC-ENS$_{16}$ outperforms SAT by 1.1 percentage points in mean accuracy and 10.6 percentage points in 5th-percentile accuracy. The benefit of ensemble training is further evident at this variability: with the same number of updates, MC-ENS$_{16}$ achieves 98.1\% mean accuracy, compared with 84.3\% for MC-ENS$_1$ (Supplemental Material, Sec.~S5). At $\eta=4$, MC-ENS$_{16}$ retains 93.7\% mean accuracy, whereas PD drops to 67.9\%.

Figure~\ref{fig:classification}(c) quantifies the fidelity of the circuit realized on an unseen chip relative to the designed circuit. The fidelity decreases from approximately 99\% at $\eta=0.5$ to 27--37\% at $\eta=4$ across all methods. Although MC-ENS$_{16}$ has the highest mean fidelity for $\eta\ge1$, its advantage for $\eta\ge2$ is 2.4--4.3 percentage points over SAT and 0.8--2.9 over $\Sigma$-SAT, compared with 9.9 percentage points over PD at $\eta=3$ (Supplemental~Material, Sec.~S9). Notably, at $\eta=3$, MC-ENS$_{16}$ achieves 98.1\% mean classification accuracy despite a circuit fidelity of only 58.7\%. This disparity indicates that accurate reproduction of the designed circuit is not necessary for reliable classification. Rather, classification is preserved when fabrication-induced deviations do not alter the predicted labels, a task-level robustness mechanism examined through the margin analysis in Fig.~\ref{fig:classification}(d).

For each sample, the mean-design margin is
$m=z_y-\max_{c\ne y}z_c$, with class scores $z_c$ and true label $y$.
With controls frozen, the same sample has margin $m+\delta m$ on chip
$\bm\xi$, where $\delta m$ is the change in the score difference.
The higher score determines the predicted class, which changes exactly
when $m(m+\delta m)<0$, apart from ties~\citep{tsuzuku2018lipschitz}. We define
$\gamma=\lvert m\rvert/s_{\delta m}$, where
$s_{\delta m}=(\E_{\bm\xi}\delta m^2)^{1/2}$ is the root-mean-square (RMS)
shift. Figure~\ref{fig:classification}(d)
pools these eight values of $\eta$ and shows higher prediction-change
probabilities at smaller $\gamma$. At $\eta=3$, the RMS shift exceeds
the nominal margin magnitude ($\gamma<1$) for 18.2\% of PD test samples
and 1.3--1.7\% for the three shared-control
methods (Supplemental~Material, Sec.~S5).

\section{Discussion and Conclusion}
\label{sec:discussion}
For small, centered fabrication perturbations, the local model
predicts that optimal device-specific compensation reduces the mean
squared operator loss on training chips and increases it by the same
amount on independent chips, relative to fixed controls that are exact
on the mean-design circuit. Shared training avoids retaining a
chip-specific correction and can reduce fabrication sensitivity, as
observed for CNOT and Ising.

MC-ENS uses prior knowledge of static chip-to-chip variability in
the form of a distribution over fabrication parameters. Inspired by
noise-aware training~\citep{joshi2020accurate,mourgias2022noise}, we
sample virtual chips from this distribution to train shared controls
for deployment on unseen chips. \SigSAT\ improves on SAT in all four
quantum tasks, and MC-ENS$_{16}$ gives the lowest mean unseen-chip errors
across the tested variability range. More importantly, the classification
diagnostics show how fabrication-induced changes in
optical outputs can leave predicted classes unchanged.
Differences in mean classification accuracy among the shared-control
methods become apparent above $\eta=2.5$, with MC-ENS$_{16}$ retaining
higher accuracy at larger variability.
In the correlated operator-learning test, misspecified fabrication
statistics reduce the transfer gains of MC-ENS and \SigSAT\
(Supplemental~Material, Sec.~S7).

Our simulations demonstrate improved cross-chip
transfer with MC-ENS. At each fabrication scale, a single control set
per task or target is reused without device-specific retraining on
unseen chips drawn from the same fabrication distribution. The local
analysis explains the transfer cost of chip-specific compensation.
The classification results illustrate how task performance can
remain stable despite changes in the optical circuit. At $\eta=3$,
MC-ENS$_{16}$ retains 98.1\% mean classification accuracy. The mean circuit
fidelity over the full computational subspace is 58.7\%, and the mean
conditional output-state fidelity for the input used by the classifier is
83.7\%. Predictions change less often when the decision margin is large
relative to the fabrication-induced margin shift.
Experiments combining measured fabrication statistics~\citep{xing2023spatial}
with in situ training~\citep{pai2023backpropagation} can quantify these
transfer gains and measurement costs on hardware.

\section*{Data and Code Availability}
All quantum-task data are generated numerically from the fabrication
model, target definitions and random-seed protocol described in
Sec.~\ref{sec:model} and the Supplemental Material, Secs.~S1 and S4.
The Banknote authentication dataset is publicly available from the UCI
Machine Learning Repository~\citep{lohweg2012banknote}. The source data
underlying the figures (SD1--SD13) are provided with the Supplemental
Material. The code used to train the models, evaluate the test chips
and generate the figures is available from the corresponding author
upon reasonable request.

\begin{acknowledgments}
We acknowledge support from the National Natural Science Foundation of China under Grant No. 62571343, Shenzhen Science and Technology Program (Grant No. ZDYJ20251211120859002), the Guangdong Provincial Quantum Science Strategic Initiative under Grant No. GDZX2503005, and Guangdong Basic and Applied Basic Research
Foundation(No. 2026A1515011661).
\end{acknowledgments}

\bibliography{references}

\end{document}